\documentclass[trackchanges,twocolumn]{aastex701}

\usepackage{amsmath,amssymb}
\usepackage{hyperref}
\usepackage{stfloats,placeins}
\usepackage{graphicx}
\usepackage{float}

\begin{document}

\title{Spectroscopic Case Studies of Four Long-Duration Transition-Region Explosive Events}

\author[0000-0002-0464-6760]{Yi'an Zhou}
\affiliation{Yunnan Observatories, Chinese Academy of Sciences, Kunming Yunnan 650216, PR China}
\affiliation{Yunnan Key Laboratory of the Solar physics and Space Science, Kunming 650216, PR China}
\email[show]{zhouyian@ynao.ac.cn}  

\author[0000-0003-2891-6267]{Xiaoli Yan}
\affiliation{Yunnan Observatories, Chinese Academy of Sciences, Kunming Yunnan 650216, PR China}
\affiliation{Yunnan Key Laboratory of the Solar physics and Space Science, Kunming 650216, PR China}
\email[show]{yanxl@ynao.ac.cn}  

\author[0000-0002-6526-5363]{Zhike Xue}
\affiliation{Yunnan Observatories, Chinese Academy of Sciences, Kunming Yunnan 650216, PR China}
\affiliation{Yunnan Key Laboratory of the Solar physics and Space Science, Kunming 650216, PR China}
\email{zkxue@ynao.ac.cn}  

\author[0000-0003-0236-2243]{Liheng Yang}
\affiliation{Yunnan Observatories, Chinese Academy of Sciences, Kunming Yunnan 650216, PR China}
\affiliation{Yunnan Key Laboratory of the Solar physics and Space Science, Kunming 650216, PR China}
\email{yangliheng@ynao.ac.cn}  

\author[0000-0003-4393-9731]{Jincheng Wang}
\affiliation{Yunnan Observatories, Chinese Academy of Sciences, Kunming Yunnan 650216, PR China}
\affiliation{Yunnan Key Laboratory of the Solar physics and Space Science, Kunming 650216, PR China}
\email{wangjincheng@ynao.ac.cn}  

\author[0000-0002-9121-9686]{Zhe Xu}
\affiliation{Yunnan Observatories, Chinese Academy of Sciences, Kunming Yunnan 650216, PR China}
\affiliation{Yunnan Key Laboratory of the Solar physics and Space Science, Kunming 650216, PR China}
\email{xuzhe6249@ynao.ac.cn}

\begin{abstract}
	This work presents a detailed spectroscopic case study of four long-duration transition-region explosive events (EEs) observed in AR NOAA 13213 on 2023 February 10 using the Interface Region Imaging Spectrograph (IRIS). The dynamic spectral evolution of each event is tracked through multi-component Gaussian fitting of the \ion{Si}{4} 1403 \AA\ line profiles. Three recurrent spectral morphologies are identified and characterized: bilateral wing enhancement, exclusive red-wing enhancement, and exclusive blue-wing enhancement, among which bilateral enhancement is the most common in the studied cases. Throughout their lifetimes of 20--25 minutes, these events display sustained and evolving bidirectional flows, with high-velocity components ($|v| > 100\ \mathrm{km\ s^{-1}}$) emerging in late phases. These spectral signatures are interpreted as evidence of ongoing or recurrent magnetic reconnection, where bilateral profiles correspond to bidirectional outflows, and exclusive wing enhancements represent geometric or evolutionary phases of the same process. In contrast, co-temporal flare ribbons and loop structures exhibit pronounced, unidirectional redshifts. This study underscores that significant non-Gaussian wing enhancement, rather than exclusively high speed, constitutes a defining spectroscopic signature of EEs, and provides detailed kinematic constraints on the dynamics of such TR explosive events.
\end{abstract}


\section{Introduction}

The resonant \ion{Si}{4} 1393.7~\AA{} and 1402.8~\AA{} lines, formed in the solar transition region, generally exhibit single-Gaussian profiles in quiet-Sun conditions. During transition-region explosive events (EEs), however, these profiles often become significantly non-Gaussian, characterized by pronounced wing enhancements with Doppler shifts reaching velocities of several tens to over $100~\mathrm{km}\mathrm{s}^{-1}$ \citep{1983ApJ...272..329B,2004ApJ...603L..57M}. Investigations to date have primarily concentrated on EEs occurring in quiet-Sun areas and coronal holes, where typical spatial extents range from $2\arcsec$ to $5\arcsec$ and durations span several tens of seconds \citep{1989SoPh..123...41D,1997Natur.386..811I,2003A&A...403..287P,2004A&A...427.1065T}. These EEs are generally observed as persistent bursts or transient brightenings with lifetimes varying from a few minutes up to tens of minutes \citep{1997SoPh..175..341I,2006A&A...446..327D}.

Explosive events have frequently been attributed to photospheric magnetic cancellation \citep{1991JGR....96.9399D,1998ApJ...497L.109C,2015ApJ...809...82G,2014ApJ...797...88H}. Numerical simulations further suggest that such features may originate from magnetic reconnection \citep{1999SoPh..185..127I,2001A&A...370..298R,2001A&A...375..228R,2015ApJ...813...86I}. In addition, deviations from Gaussian spectral profiles observed in explosive events may reflect dynamic processes such as spinning, unwinding, or twisting motions \citep{2011A&A...532L...9C,2012SoPh..280..417C}. For instance, \citet{2014Sci...346D.315D} showed that small-scale transition region loops exhibiting twisting motions can generate spectral signatures consistent with explosive events. Moreover, \citet{2014ApJ...797...88H} reported associations with plasma ejections followed by chromospheric retraction. 

{The physical origin of transition-region EEs remains debated, as their spectral signatures may reflect different scenario.}
Observations have linked EEs to spicular and macrospicular activity in the chromosphere \citep{2000A&A...360..351W} and to chromospheric upflow events \citep{1998ApJ...504L.123C}. In small‐scale loops, EEs have been interpreted as manifestations of siphon flows \citep{2004A&A...427.1065T}, and \citet{2009ApJ...701..253M} proposed that upflows and downflows within surges could produce the characteristic non‐Gaussian line profiles. {Here, ‘surges’ refer to jet-like chromospheric ejections typically seen as dark or bright linear features in H$\alpha$, commonly linked to magnetic reconnection events.} Explosive events have also been detected in conjunction with transient EUV brightenings and X‐ray jets \citep{2012A&A...545A..67M}. A statistical investigation by \citet{2019ApJ...873...79C} further indicates that, while a subset of EEs coincides with the onset or evolution of network jets, many events exhibit no clear association.

EEs predominantly appear in transition‐region emission lines formed at temperatures between $2\times10^{4}$ and $5\times10^{5}\,\mathrm{K}$ \citep{1983ApJ...272..329B}, including \ion{Si}{4} 1394/1403\,\AA, \ion{O}{4} 1032\,\AA, and \ion{C}{4} 1548/1550\,\AA{} lines. High‐Resolution Telescope and Spectrograph (HRTS) observations indicate that fewer than 1\% of EEs detected in these transition‐region lines have counterparts in the cooler \ion{C}{1} 1561\,\AA\ line (formation temperature $1\times10^{4}\,\mathrm{K}$), with only marginal signatures in \ion{C}{2} 1335\,\AA\ ($1.6\times10^{4}\,\mathrm{K}$) \citep{1991JGR....96.9399D}. Solar Ultraviolet Measurements of Emitted Radiation (SUMER) data further demonstrate that EEs can manifest in lower‐temperature transitions such as \ion{O}{1} ($1\times10^{4}\,\mathrm{K}$), Lyman‐6 through Lyman‐11 ($1.2\times10^{4}\,\mathrm{K}$; \citep{2002A&A...382..319M}), and Lyman-$\beta$ ($1.2\times10^{4}\,\mathrm{K}$; \cite{2010A&A...520A..37Z}). These cooler lines typically display non‐Gaussian profiles characterized by a central reversal flanked by dual emission peaks. In particular, Lyman-$\beta$ profiles during EEs exhibit enhanced self‐absorption and a pronounced intensity excess in the blue wing \citep{2010A&A...520A..37Z}.

With the help of simultaneous imaging and spectroscopic observations from the Interface Region Imaging Spectrograph (IRIS; \citep{2014SoPh..289.2733D}), we identify explosive events occurring in quiet-Sun regions. These EEs have lifetimes of 20--25~minutes and exhibit significant \ion{Si}{4} intensity enhancements relative to the surrounding quiet-Sun background. Following the event-selection criteria of \citet{2017MNRAS.464.1753H}, we extracted and line-fit the spectral profiles of 4 EEs. The manuscript is organized as follows: Section~\ref{obs and data} outlines the observations and data reduction procedures; Section~\ref{res} presents the spectral fitting results for representative EEs; and Section~\ref{conclu and discuz} discusses the implications of our findings and summarizes the conclusions.

\section{Observations and data reduction}
\label{obs and data}

\begin{figure*}[htbp]
	\centering
	\includegraphics[width=\hsize]{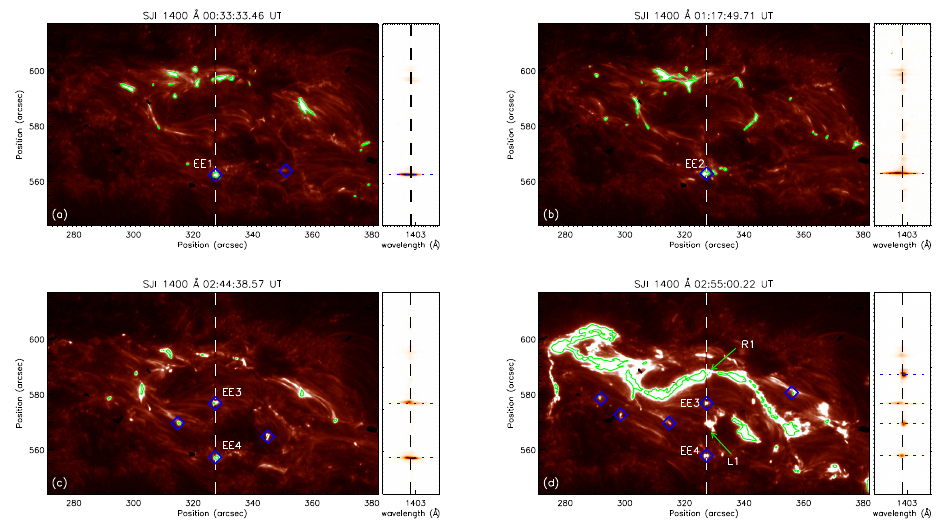}
	\caption{Overview of transition region explosive events (EEs) in AR NOAA 13213, observed with IRIS slit-jaw images (SJIs) at 1400~\AA\ and the corresponding \ion{Si}{4} 1403~\AA\ spectra near 00:33:33~UT, 01:17:50~UT, 02:44:39~UT, and 02:55:00~UT. In each SJI panel, the vertical white dashed line indicates the slit position. EE1--EE4 and the blue diamond denote the locations of EEs. Green contours outline regions with intensities exceeding 5\% of the maximum within the field of view. In the spectral panels, vertical black dashed lines mark the line center of \ion{Si}{4} 1403~\AA, while horizontal black dashed lines indicate the positions of EEs, flare ribbons, and loop structures.}
	\label{fig:sji}
\end{figure*}

\begin{figure*}[htbp]
	\centering
	\includegraphics[width=\hsize]{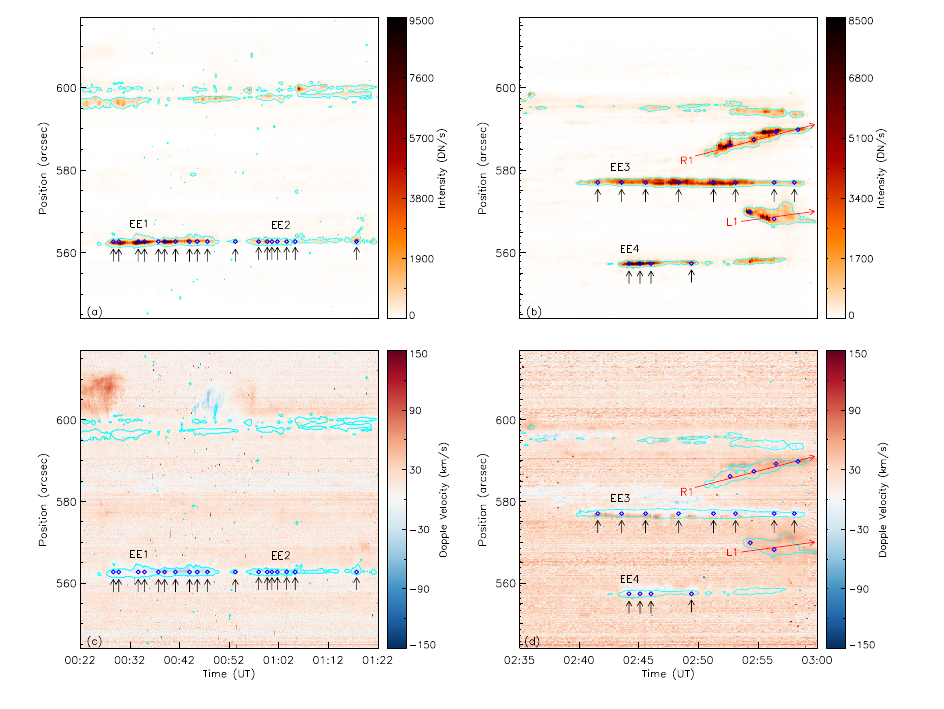}
	\caption{Panels (a) and (b): space–time maps of the total intensity of the \ion{Si}{4} 1403~\AA\ line and the corresponding intensity of explosive events (EEs), flare ribbons, and loop structures. Cyan contours delineate areas with an intensity 0.05 times the maximum intensity of observed region. Specific locations of EEs, flare ribbons, and loops selected for further analysis are marked by blue diamonds and black arrows.
	Panels (c) and (d): same as (a) and (b), but for the Doppler velocities.}
	\label{fig:map}
\end{figure*}

We analyze several transition-region (TR) explosive events (EEs) observed by the IRIS in AR NOAA 13213 on 2023 February 10. IRIS acquired high-cadence sit-and-stare spectral observations during two intervals: 00:10:19–01:22:22~UT and 01:46:33–03:00:29 UT. The spectroscopic data exhibit a temporal resolution of $\sim$9.3~s with a slit width of $\sim$0\farcs33. For the \ion{Si}{4} far-ultraviolet emission lines, the spectral resolution is $\sim$0.025~\AA. Complementary 1400~\AA{} slit-jaw images (SJIs), obtained with a $\sim$28 s cadence, reveal multiple TR EEs during these intervals (Figure~\ref{fig:sji}), which we demarcate with blue diamond symbols. EEs intersected by the IRIS slit are designated EE\textsubscript{N} (where N denotes the event identifier), with corresponding spectra presented at their respective observation times (EE1-EE4 in Figure~\ref{fig:sji}(a)-(d)). Notably, distinct flare ribbons and loop structures were observed during the second observational period (Figure~\ref{fig:sji}(d)). 

To investigate the evolution of EEs, flare ribbons, and loop structures along the IRIS slit, we constructed time-space diagrams of the Si iv 1403~\AA{} integrated intensity for two observational intervals (Figure \ref{fig:map}). Cyan contours delineate areas with an intensity 0.05 times the maximum intensity of observed region. Arrows mark spatial positions corresponding to the EEs, flare ribbons, and loop structures identified in Figure~\ref{fig:sji}. 
Furthermore, in computing the \ion{Si}{4} 1403~\AA{} integrated intensity, spatial averaging was applied over $\pm$3 pixels centered on each target position with subtraction of the mean intensity from a background quiet region.

As shown in panel (a) of Figure~\ref{fig:map}, a distinct explosive event intersected by the IRIS slit during the first observational interval is located at approximately 562\arcsec--563\arcsec. This event commenced rapid brightening around 00:26~UT, attained peak intensity near 00:30~UT and 00:34~UT (designated EE1), and exhibited multiple quasi-periodic brightenings over the subsequent 10 minutes until initial dimming occurred at 00:48~UT. Subsequently, it underwent renewed intensification around 00:54~UT, reaching secondary intensity maxima near 00:58~UT and 01:18~UT (EE2), followed by gradual decay.

During the second observational period, two distinct explosive events were detected at approximately 557\arcsec and 575\arcsec (panel (b) of Figure~\ref{fig:map}). The northern event (EE3) commenced significant brightening around 02:39~UT, persisted for approximately 20 minutes, and initiated rapid decay near 02:59~UT. The southern event exhibited delayed onset, undergoing rapid intensification from 02:43~UT and attaining peak intensity near 02:45~UT (EE4). This event maintained enhanced brightness for nearly 10 minutes before gradual decay commenced.

Additionally, the flare ribbons (panel (b) of Figure~\ref{fig:map}) demonstrated rapid brightening onset around 02:52~UT, achieved maximum intensity within 5 minutes, and subsequently underwent rapid dimming. As evident in panel (b) of Figure~\ref{fig:map}, the flare ribbon system exhibited systematic northward displacement. Furthermore, a transient loop structures detected at the intermediate position ($\sim$570\arcsec) between the two EEs displayed rapid brightening commencement near 02:53~UT, attained peak intensity around 02:55~UT, and subsequently decayed.

Subsequently, we derived the space-time distribution of Doppler velocities for the \ion{Si}{4}~1403~\AA{} emission line using the moment method (Figure~\ref{fig:map}{(c) and (d)}). The centroid of \ion{Si}{4} line was calibrated against the photospheric \ion{S}{1}~1401.5~\AA{} reference line within the same spectral window, yielding a rest wavelength of $1402.76$~\AA{}. For velocity derivation, the spectral sampling was constrained to regions where the \ion{Si}{4} intensity exceeded $3\sigma$ above the far line-wing intensity baseline. 
{We then removed saturated pixels and performed deblending to isolate the \ion{Si}{4}~1403~\AA{} line from blended spectral features. This process involved identifying and masking pixels contaminated by nearby emission lines (e.g., from cooler species or fixed-pattern noise) using a reference quiet-Sun spectrum obtained from the same observational sequence. A local continuum was then fitted and subtracted in the spectral region surrounding the \ion{Si}{4} line to remove any residual blended background. Finally, the background intensity, quantified as the mean intensity from 20 quiescent slit positions at each time step, was subtracted as a constant offset from each profile to yield the clean \ion{Si}{4}~1403~\AA{} profile for analysis.}

The observations (Figure~\ref{fig:map}) reveal that explosive events EE1--EE4 exhibit both net red/blue-shifts with Doppler velocities ranging from -70--$50~\mathrm{km}~\mathrm{s}^{-1}$. Conversely, the flare ribbons and loop structure display dominant redshift signatures with velocities of 10--$70~\mathrm{km}~\mathrm{s}^{-1}$. To characterize their spectral properties, we performed multi-component Gaussian fits on individual \ion{Si}{4} line profiles, deriving velocity and broadening parameters for distinct spectral components.

\section{Results}
\label{res}
We performed Gaussian fitting and statistical analysis of \ion{Si}{4}\,1403\,\AA{} line profiles spanning four event intervals: EE1 (00:25:33--00:50:03~UT), EE2 (00:55:04--01:22:13~UT), EE3 (02:39:47--03:00:02~UT), and EE4 (02:41:39--02:58:27~UT).  
{To characterize the dynamic spectral profiles, we performed multi-component Gaussian fitting on the \ion{Si}{4}~1403~\AA{} line. The fitting strategy was conservative: we started with a single Gaussian and added additional components only when they significantly reduced the fit residuals (assessed via the reduced $\chi^{2}$ statistic) and corresponded to visually distinct emission enhancements in the line wings (e.g., secondary peaks or pronounced asymmetries). Most profiles during the explosive events were adequately represented by two or three components: a near‑core component and one or two Doppler‑shifted wing components. The uncertainties of the fitted parameters (intensity, centroid, width) were derived from the covariance matrix, and only components with amplitudes exceeding 3 times of their uncertainty were retained. Although non‑thermal broadening may contribute to the line width, the clear discrete Doppler shifts in the wings support the use of Gaussian components as an empirical decomposition to track the dominant flow structures.}

\begin{figure*}[htbp]
	\centering
	\includegraphics[width=\hsize]{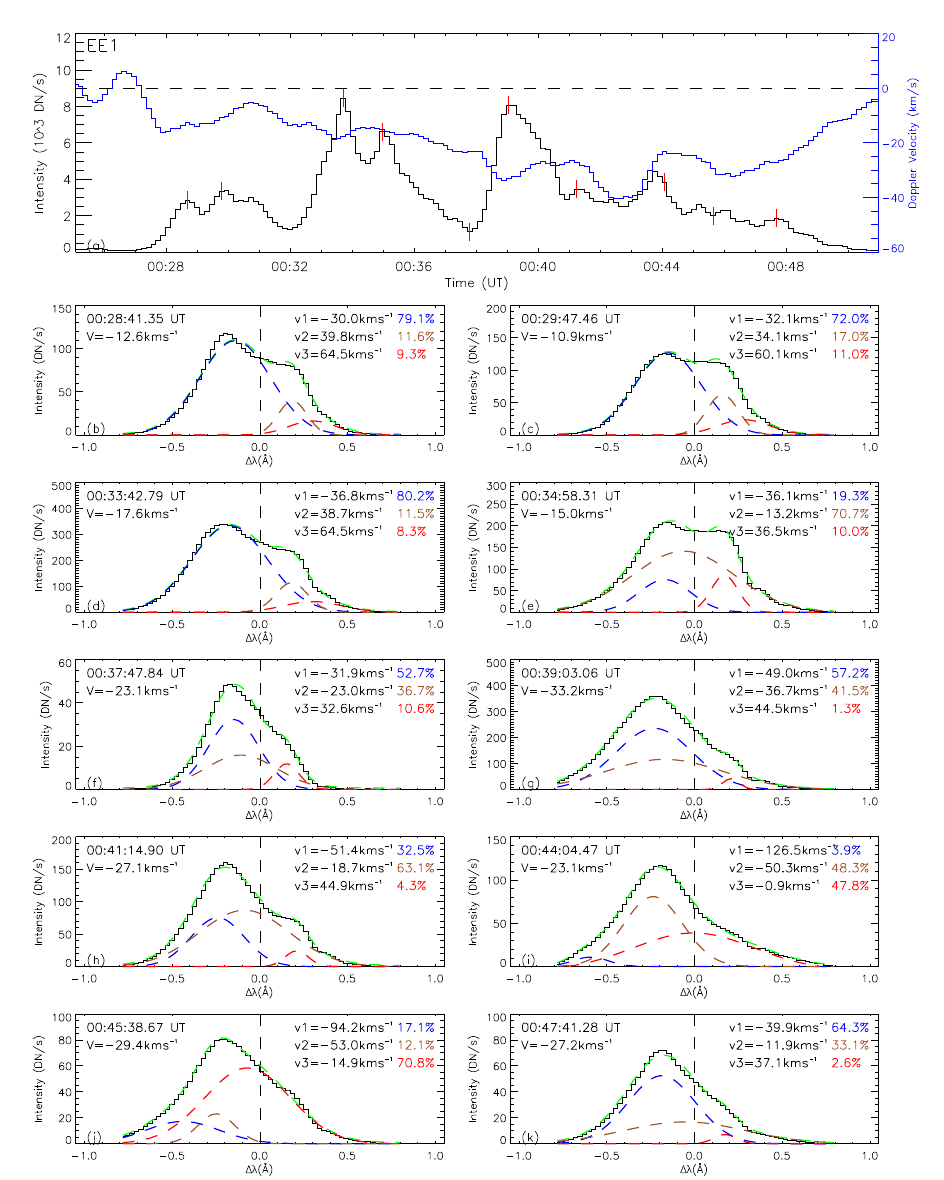}
	\caption{Panel (a): time evolution of the integrated intensity (black curve) and Doppler velocity (blue curve) of \ion{Si}{4} 1403~\AA\ line. The short vertical red lines indicate the specific position marked by blue diamonds in Figure~\ref{fig:map}. 
	Panel (b)--(k): the \ion{Si}{4} 1403~\AA\ line profiles of EE1. In each panel, the vertical dashed lines mark the rest wavelength of the \ion{Si}{4} 1403~\AA\ line. Green dashed curves indicate the composite Gaussian fits, while blue, red, and orange curves denote the individual Gaussian components. The global velocity, $v$, is derived from the centroid of the fitted profile. The velocities $v_{1}$, $v_{2}$, and $v_{3}$ correspond to the Doppler shifts of each Gaussian component.
	}
	\label{fig:profile1}
\end{figure*}

\begin{figure*}[htbp]
	\centering
	\includegraphics[width=\hsize]{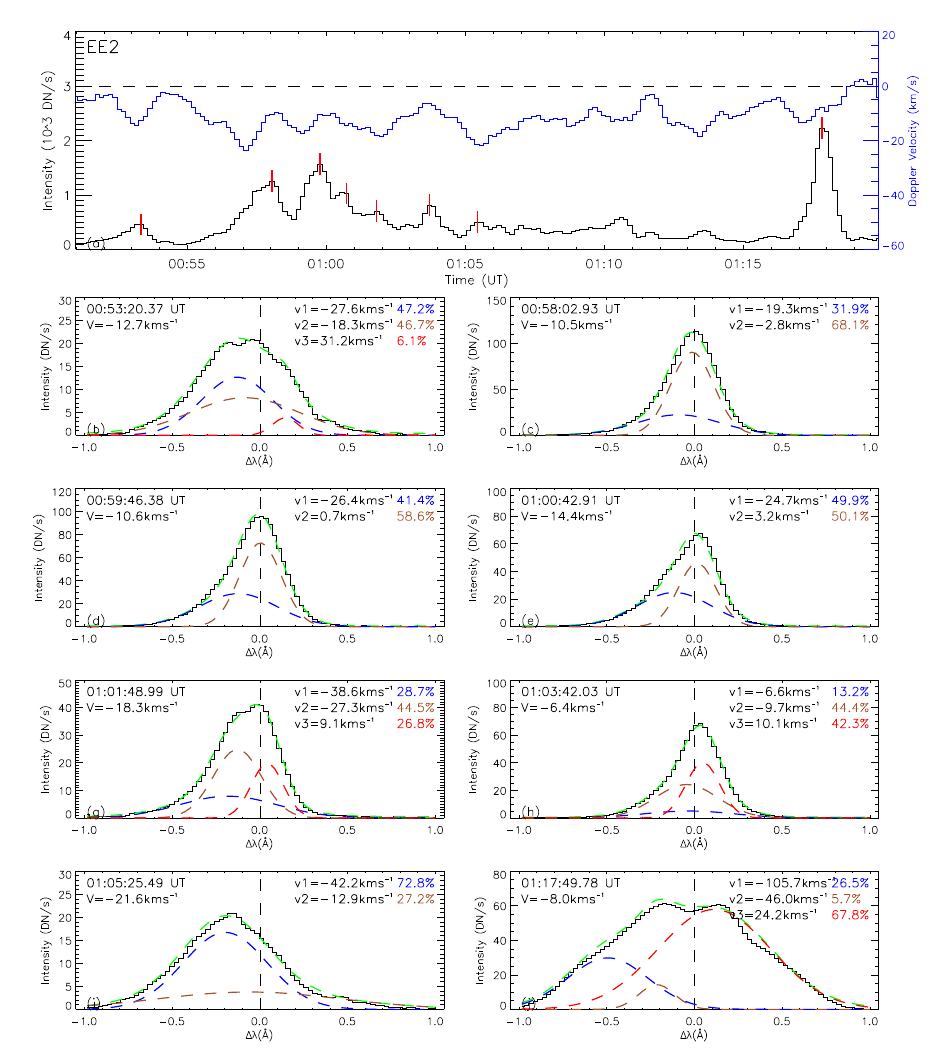}
	\caption{Same as Figure \ref{fig:profile1}, but for the line profiles of EE2.
	}
	\label{fig:profile2}
\end{figure*}

\begin{figure*}[htbp]
	\centering
	\includegraphics[width=\hsize]{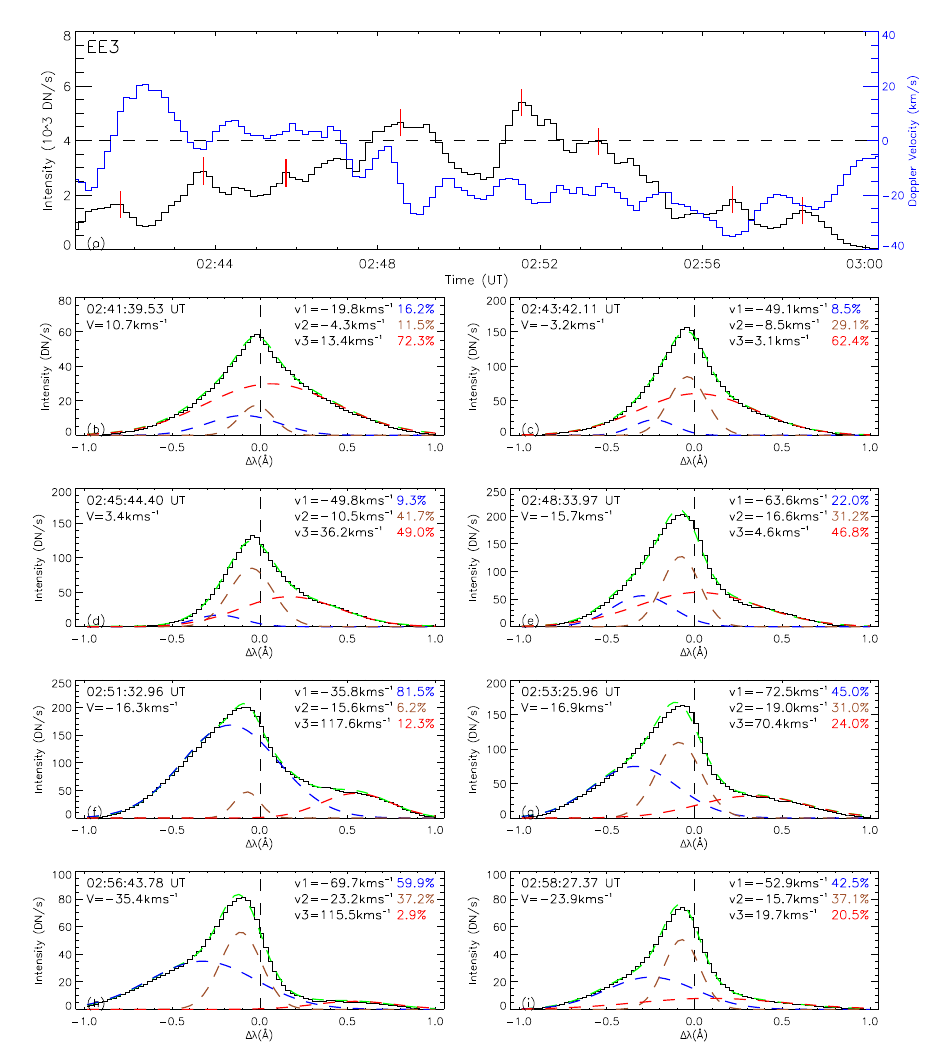}
	\caption{Same as Figure \ref{fig:profile1}, but for the line profiles of EE3.}
	\label{fig:profile3}
\end{figure*}

\begin{figure*}[htbp]
	\centering
	\includegraphics[width=\hsize]{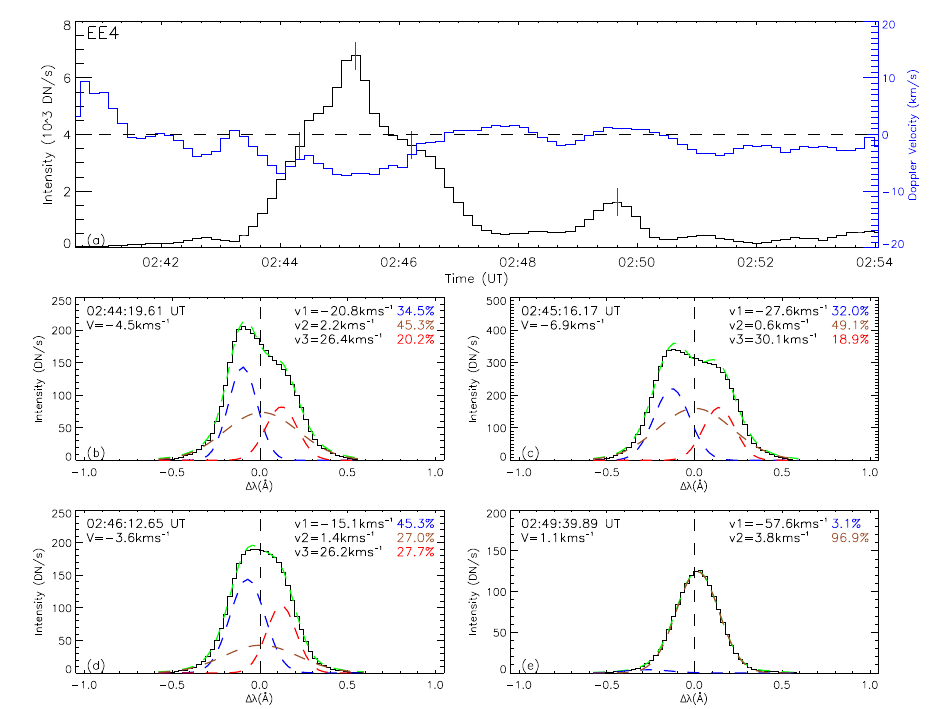}
	\caption{Same as Figure \ref{fig:profile1}, but for the line profiles of EE4.}
	\label{fig:profile4}
\end{figure*}

{The analysis demonstrates that while individual explosive events may display totally red or blue shifts, their spectral line profiles can nonetheless display three distinct morphologies: (1) bilateral enhancement in both wings, (2) enhancement exclusively in the red wing, and (3) enhancement exclusively in the blue wing. The morphology characterized by bilateral enhancement in both wings is the dominant one. Representative \ion{Si}{4}~1403~\AA{} line profiles in EEs for these features, corresponding to blue diamond markers in Figure~\ref{fig:map}, are presented in Figures~\ref{fig:profile1}--\ref{fig:profile4}. Besides, the temporal evolution of the \ion{Si}{4} 1403~\AA{} intensity and Doppler velocities of EE1 to EE4 is presented in panel (a) of Figures~\ref{fig:profile1}--\ref{fig:profile4}. }

\begin{figure*}[htbp]
	\centering
	\includegraphics[width=\hsize]{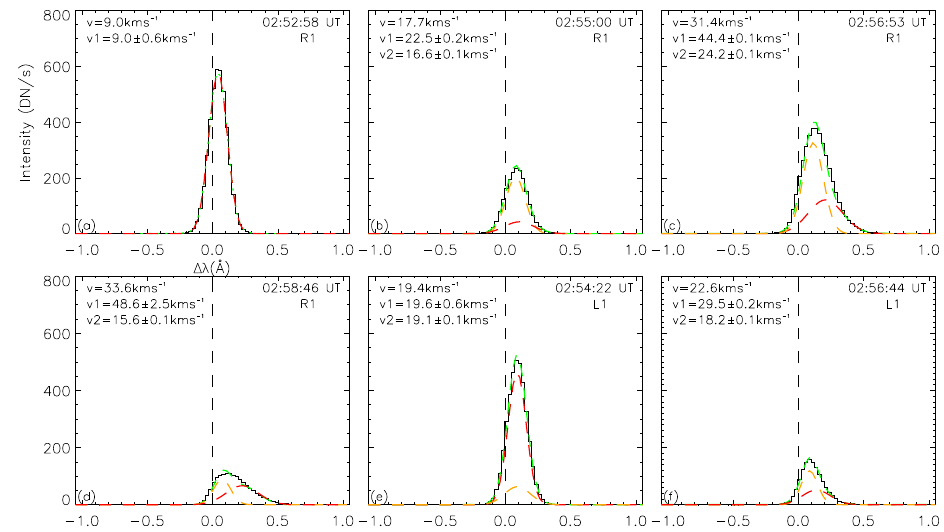}
	\caption{Same as Figure \ref{fig:profile1}, but for the line profiles at flare ribbons and loop structures.}
	\label{fig:rl}
\end{figure*}

In addition, both flare ribbons (R1) and loop structures (L1) display globally redshifted profiles with pronounced red-wing enhancement. {Figure \ref{fig:rl} displays the spectral line profiles and the corresponding Gaussian fits.}

\subsection{\ion{Si}{4} line profiles of EE1}

Figure~\ref{fig:profile1}{(a)} reveals that explosive event EE1 predominantly exhibits net blueshifts, with velocities ranging from several to tens of $\mathrm{km}~\mathrm{s}^{-1}$.
{Throughout its evolution, the intensity of EE1 experienced multiple enhancements, manifesting as several distinct peaks.
The representative \ion{Si}{4} 1403~\AA\ line profiles indicated by blue diamonds in Figure~\ref{fig:map} are marked by short red vertical ticks in Figures~\ref{fig:profile1}{(a)}.}
These profiles were then decomposed via multi-component Gaussian fitting, with blue- and redshifted components represented by blue and red dashed lines respectively, and near-core components indicated by brown dashed lines (Figures~\ref{fig:profile1}{(b)--(k)}).

{The evolution of EE1 is characterized by persistent bilateral wing enhancement and significant line broadening throughout the interval (Figures~\ref{fig:profile1}{(b)--(k)}). Based on its intensity evolution, the event can be divided into three distinct phases: 00:26--00:32~UT, 00:32--00:38~UT, and 00:38--00:50~UT.}

{During the first enhancement interval (Figures~\ref{fig:profile1}{(b) and (c)}), the \ion{Si}{4} line profiles at the two intensity maxima exhibit clear enhancements in both the red and blue wings. The blue-wing component dominates the total intensity, contributing more than 70\%. 
The second enhancement is significantly stronger than the first (Figures~\ref{fig:profile1}{(d)--(f)}), reaching its peak near 00:34~UT. As in the earlier interval, a blue-wing enhancement at a Doppler shift of $\sim$30 $\mathrm{km}~\mathrm{s}^{-1}$ persists, while the velocity of the red-wing enhancement gradually decreases. Throughout this period, the fractional intensity contributed by near-core component increases noticeably.
Approximately one minute after the second enhancement, EE1 intensifies rapidly reached its third peak near 00:39~UT, attaining a peak intensity comparable to that of the second interval, and after which EE1 decays and disappears (Figures~\ref{fig:profile1}{(g)--(k)}). In this final phase, the near-core component becomes the dominant contributor to the total intensity, while the relative contribution from the blue-wing enhancement diminishes with time. Toward the late stages of the event, the fractional intensity of the red-wing enhancement increases (Figures~\ref{fig:profile1}{(i) and (j)}). In addition, a notable feature during the third interval is a sharp increase in the Doppler velocity of the blue-wing component, from several tens of $\mathrm{km}~\mathrm{s}^{-1}$ to approximately 100 $\mathrm{km}~\mathrm{s}^{-1}$.}

\subsection{\ion{Si}{4} line profiles of EE2}
{Explosive event EE2 exhibits comparatively weaker line intensity and smaller net blueshift than EE1 (Figure~\ref{fig:profile2}{(a)}). Its decomposition shows a lower-velocity component, with instances of exclusive blue-wing enhancement and bilateral wing enhancement (Figure~\ref{fig:profile2}{(b)--(j)}).}

{Similar to EE1, the evolution of EE2 can be divided into two principal intervals: 00:51--01:15~UT and 01:15--01:20~UT. During the first pronounced enhancement (Figure~\ref{fig:profile2}{(b)--(i)}), the total intensity is dominated by the component near the line core, with the blue-wing enhancement providing a secondary contribution. At several peaks near 01:00~UT, the \ion{Si}{4} line profiles exhibit exclusively blue-wing enhancement (Figure~\ref{fig:profile2}{(c)--(e)}) and are well fitted by a two-Gaussian fit. In these cases, the Doppler velocity of the blue-wing component is $\sim$20 $\mathrm{km}~\mathrm{s}^{-1}$, while the near-core component is confined to within a few $\mathrm{km}~\mathrm{s}^{-1}$ of the rest wavelength.}
	
{As EE2 evolves, the blue-wing enhancement gradually weakens and a distinct red-wing enhancement emerges (Figure~\ref{fig:profile2}{(g)--(h)}), disappearing a few minutes later (Figure~\ref{fig:profile2}{(i)}). The red-wing components exhibit Doppler velocities on the order of $\sim$10 $\mathrm{km}~\mathrm{s}^{-1}$, and their fractional intensity increases until it becomes comparable to that of the near-core component. During the second enhancement phase, the line width increases noticeably and a clear simultaneous enhancement of both wings is again observed; at this stage the Doppler velocity of the blue-wing component reaches $\sim$100 $\mathrm{km}~\mathrm{s}^{-1}$.}

\subsection{\ion{Si}{4} line profiles of EE3}

{Explosive event EE3 exhibits distinct characteristics relative to EE1 (Figure~\ref{fig:profile3}). Their \ion{Si}{4} line profiles display significant broadening with comparatively lower intensity than that of EE1. Additionally, these demonstrate both net redshifts and blueshifts.
Gaussian fits reveal a persistent low-velocity component near the line core in EE3, with velocities of several to tens of $\mathrm{km}~\mathrm{s}^{-1}$ (brown dashed lines in Figure~\ref{fig:profile3}{(b)--(i)}), irrespective of the global profile morphology.
Similar to EE1 and EE2, the intensity evolution of EE3 can be separated into two distinct intervals: 02:41--02:51~UT and 02:51--03:00~UT, each lasting approximately 10 minutes. The \ion{Si}{4} line profiles are consistently well described by simultaneous bilateral wing enhancement.}

{During the first enhancement interval, the integrated line intensity is dominated by the red-wing enhancement component, with the component near the line core providing the secondary contribution. As the total intensity of EE3 gradually increases, the fractional contribution from the red wing decreases while that from the blue wing increases (Figure~\ref{fig:profile3}{(b)--(e)}). The Doppler velocities of the red-wing components range from a few to several tens of $\mathrm{km}~\mathrm{s}^{-1}$, while the blue-wing components exhibit velocities of order several tens of $\mathrm{km}~\mathrm{s}^{-1}$.}

{EE3 attains its peak intensity near 02:52~UT during the second enhancement interval, after which the total intensity gradually declines. This phase exhibits significantly stronger spectral line broadening than the first, with the blue-wing enhancement now dominating the total intensity. As EE3 fades, the fractional contribution of the blue wing decreases, while the contribution from the line-core component remains relatively stable. Notably, during this second enhancement the Doppler velocity of the red-wing components increases sharply, reaching up to $\sim$100 $\mathrm{km}~\mathrm{s}^{-1}$, yet the overall line profile maintains a net blueshift.}

\subsection{\ion{Si}{4} line profiles of EE4}
{In contrast to EE1--EE3, the temporal intensity profile of EE4 features only two distinct peaks (Figure~\ref{fig:profile4}{(a)}), and its overall Doppler shifts remain confined to within $\sim$10~$\mathrm{km~s^{-1}}$. Despite exhibiting less spectral broadening than EE1--EE3, EE4 similarly displays significant bilateral wing enhancement throughout its evolution (Figures~\ref{fig:profile4}{(b)--(d)}).}

{The line profiles are characterized most notably by a quasi-static component located very close to the line core (Figure~\ref{fig:profile4}{(b)--(d)}), exhibiting velocities of only a few $\mathrm{km~s^{-1}}$. The velocities of the corresponding red- and blue-wing enhancements remain relatively stable throughout the evolution of EE4, on the order of 10 to tens of $\mathrm{km~s^{-1}}$. By the time of the second intensity peak, the \ion{Si}{4} line profile closely approximates a single Gaussian shape typical of the quiet background, retaining only a faint enhancement in the blue wing.}

\subsection{\ion{Si}{4} line profiles of R1 and L1}

Figures~\ref{fig:map} and~\ref{fig:rl} reveal that flare ribbons and loop structures exhibit enhanced line intensities relative to explosive events, displaying net redshifts ranging from a few to several tens of $\mathrm{km}~\mathrm{s}^{-1}$. Unlike explosive events, these features predominantly show pronounced red asymmetry in their \ion{Si}{4} line profiles and significantly reduced line broadening.

Gaussian fits demonstrate that the flare ribbon profile at 02:53~UT is well characterized by a single-component fit (Figure~\ref{fig:rl}{(a)}). During its evolution, the \ion{Si}{4}~1403~\AA{} line exhibits an initial increase followed by a decrease in net redshifted magnitude, concurrent with progressively intensifying red-wing enhancement (Figures~\ref{fig:rl}{(b)--(d)}). Conversely, loop structures display measurable red-blue spectral asymmetry in their profiles (Figures~\ref{fig:rl}{(e) and (f)}).

\section{Conclusions and discussions}
\label{conclu and discuz}
{This study presents a detailed spectroscopic analysis of four long-duration explosive events (EE1–EE4), loops and flare ribbons. To characterize their dynamic evolution, we performed a multi-component Gaussian decomposition on the Si IV 1403 \AA\ line profiles across the lifetime of each explosive event. The fitting strategy was conservative: starting with a single Gaussian, additional components were introduced only when they significantly reduced the fit residuals and corresponded to visually distinct emission enhancements in the line wings (e.g., secondary peaks or pronounced asymmetries). A constant background, determined from nearby quiet regions along the slit, was subtracted from each profile to isolate the EE emission. This approach allowed us to track the temporal evolution of distinct kinematic components (e.g., near-core, red-wing, blue-wing) within each EE. The evolving profiles across all four EEs can be categorized into three morphological types: bilateral wing enhancement, exclusive red-wing enhancement, and exclusive blue-wing enhancement. The detailed evolution of each EE, as detailed below, reveals how these morphologies appear and transition over time, providing direct insight into the underlying physical processes. We interpret these observations in the context of magnetic reconnection and related flows.}

\subsection{EE1: Sustained bidirectional outflows with evolving dominance}

{The evolution of EE1 is marked by persistent bilateral wing enhancements and significant line broadening throughout its $\sim$25-minute lifetime (Figure~\ref{fig:profile1}). This dominant morphology is classically interpreted as the signature of bi-directional plasma outflows from a magnetic reconnection site\citep{2009ApJ...702....1H, 2011A&A...535A..95D,2015ApJ...813...86I,2022NatCo..13..640Y}. The event’s intensity evolution delineates three phases. In the first two phases, the blue-wing component dominates, contributing $>$70\% of the intensity with steady velocities around 30 $\mathrm{km}~\mathrm{s}^{-1}$, while a weaker, decelerating red-wing component is also present. This suggests a reconnection geometry favoring upward-directed outflows along the line of sight. A striking shift occurs in the third phase: as the overall intensity decays, the Doppler velocity of the blue-wing component abruptly increases to $\sim$100 $\mathrm{km}~\mathrm{s}^{-1}$. Concurrently, the fractional contribution from the near-core component grows, and the red-wing enhancement strengthens in the final minutes.}

{This temporal sequence—from dominant moderate-speed outflows to the sudden acceleration of one outflow channel alongside increased stationary emission—carries physical implications. The late-phase high-velocity blueshift could indicate a change in the reconnection environment, such as a shift to a lower-density region or increased reconnection rate. The growing near-core and red-wing contributions may trace the subsequent cooling and downward draining (siphon flows) of the heated plasma, a common aftermath of impulsive heating\citep{2004A&A...427.1065T}. This behavior corresponds to magnetic reconnection processes documented in quiet-Sun regions\citep{2017ApJ...845...16N,2018MNRAS.479.3274S}. Thus, EE1 may exemplify how a single reconnection site can produce a complex, evolving spectrum as the balance between acceleration, heating, and cooling processes changes over time.}

\subsection{EE2: Morphological transition from unidirectional to bidirectional flows}

{EE2 exhibits lower overall intensity than EE1 but reveals a clear morphological evolution (Figure~\ref{fig:profile2}). Its early phase is characterized by exclusive blue-wing enhancement at peaks near 01:00 UT, well-fitted by two components: a near-stationary core and a blueshifted ($\sim$20 $\mathrm{km}~\mathrm{s}^{-1}$) component. This unidirectional signature is reminiscent of reconnection-driven upflows\citep{1998ApJ...504L.123C,2009ApJ...701..253M}. As the event progresses, a distinct red-wing enhancement emerges and strengthens until its contribution rivals the core’s, transforming the profile into a bilateral type before fading. A final, brief intensification reignites clear bilateral enhancement with the blue-wing velocity reaching $\sim$100 $\mathrm{km}~\mathrm{s}^{-1}$.}

{Although such isolated enhancements have been associated with network jets \citep{2014Sci...346A.315T,2019ApJ...873...79C}, our SJI observations reveal no corresponding jet structures. The transition from exclusive blue-wing to bilateral enhancement in EE2 may suggest a change in the visibility or geometry of the reconnection outflows.  Initially, the reconnection outflow might be directed predominantly upward and aligned with the line of sight, with any downward-directed flow possibly obscured or too faint. Later, as the reconnection site evolves, the downward-directed flow becomes more pronounced or enters the spectrometer's line-of-sight, revealing the classic bidirectional signature. This morphological fluidity within a single EE underscores that the observed line profile is a sensitive function of the observer's perspective relative to the reconnection plane and the temporal evolution of the outflow structure.}

{Besides, the initial exclusive blue-wing enhancement in EE2 may correspond to reconnection within a highly asymmetric or inclined current sheet\citep{2017MNRAS.464.1753H}. In such a configuration, the upward-directed (blueshifted) outflow could be channeled along open or favorably oriented field lines, appearing bright and well-defined. The downward-directed (redshifted) outflow, however, might be injected into a diffuse, low-density, or cooler loop structure, rendering its emission too faint for detection. As the reconnection evolves, the magnetic topology may change. The reconnection site could shift to a region with more symmetric field lines, allowing the downward outflow to energize a denser or more efficiently heated loop, causing the red-wing component to emerge and intensify.}

\subsection{EE3: Different flow dominance within persistent bidirectional enhancement}
{EE3 presents a compelling case of persistent bilateral enhancement where the dominant flow component reverses (Figure~\ref{fig:profile3}). Throughout its two $\sim$10-minute phases, the profiles consistently show two-wing enhancements. Initially, the red-wing component dominates the integrated intensity. As the total intensity rises, the blue-wing's fractional contribution grows at the expense of the red wing’s, eventually dominating at the peak near 02:52 UT. Notably, during this second phase, the Doppler velocity of the red-wing components sharply increases to $\sim$100 $\mathrm{km}~\mathrm{s}^{-1}$, even though the overall profile maintains a net blueshift.}

{This behavior highlights the inherent complexity in interpreting bidirectional flows. The observed reversal in dominance, coupled with the late-stage high-speed red component, strongly suggests a dynamic physical change at the source. This could involve a spatial shift of the reconnection site, altering the projection and radiative contribution of each outflow, or an evolution in the local plasma conditions (density, temperature) affecting their emissivity\citep{2024MNRAS.529.3424H}. The persistence of a net blueshift alongside a high-velocity red component underscores that multiple, independently evolving flows coexist within the observational resolution element. The profile is a dynamic intensity-weighted average, not a static structure. An alternative interpretation within the rotating jet framework \citep{2012SoPh..280..417C} posits that such spectral evolution could also arise from changes in the dominant rotational component or the development of localized high-velocity twists within a single coherent structure.}

\subsection{EE4: Weak flows and the role of a quasi-static Core}

{EE4 stands in contrast to the others, showing minimal net Doppler shifts ($<$10 $\mathrm{km}~\mathrm{s}^{-1}$) and only two intensity peaks, yet still displaying clear bilateral wing enhancements (Figure~\ref{fig:profile4}). Its most distinctive feature is a strong, quasi-static component at the line core (velocities of a few $\mathrm{km}~\mathrm{s}^{-1}$), accompanied by stable, moderate-velocity ($\sim$10--40 $\mathrm{km}~\mathrm{s}^{-1}$) wing enhancements. By its second peak, the profile converges to nearly a single Gaussian.}

{The characteristics of EE4 suggest it may correspond to a magnetic reconnection event with intrinsically lower outflow velocities, consistent with the range of speeds observed in some small-scale reconnection events (e.g., \cite{2021ApJ...915...17X}). The strong, near-stationary core component likely represents heated plasma confined near the reconnection site. This feature finds a plausible physical explanation in simulations of reconnection in partially ionized plasma, where processes like ion-neutral coupling can lead to enhanced emission and plasma accumulation in the diffusion region \citep{2012ApJ...760..109L}. The persistent bilateral wings, even at low velocities, remain indicative of oppositely directed outflows from reconnection.}

\subsection{Flare ribbons and loop structures: unidirectional downflows}
{The spectra from the co-temporal flare ribbon (R1) and post-flare loop (L1) provide a contrasting benchmark (Figure~\ref{fig:rl}). These structures show stronger intensities, pronounced and persistent red asymmetries or net redshifts (10--70 $\mathrm{km}~\mathrm{s}^{-1}$), and significantly less broadening than the EEs. The profile in flare ribbon is often well-fit by a single redshifted Gaussian, a hallmark of chromospheric condensation---a downflow of cooled, dense plasma following explosive coronal heating\citep{2015ApJ...811..139T,2020ApJ...896..154Y}. This stark difference underscores a key finding: while the EEs exhibit bidirectional or mixed flows indicative of localized magnetic reconnection, the flare-related features are dominated by unidirectional downflows from impulsive heating and subsequent cooling. This contrast reinforces the interpretation of EEs as sites of ongoing reconnection, distinct from the hydrodynamic response patterns of flares.}

\subsection{Synthesis: The feature of the observed EEs}
{By tracing the detailed spectral evolution of four long-duration EEs, this study provides a dynamic view beyond a static morphological census. The key finding is that bilateral wing enhancement is not only a common morphology but also a persistent and evolving state within these events. To explain these observations, magnetic reconnection—particularly in a sustained or recurrent form—emerges as a primary candidate mechanism, as it naturally accounts for the persistent and evolving bidirectional flows. 
Furthermore, the height of the reconnection site relative to the transition region can influence whether one outflow component dominates in the observed profiles. Reconnection occurring well above the transition region may primarily manifest as downflow (red-wing) signatures (e.g., \cite{2015ApJ...813...86I}; \cite{2021ApJ...915...17X}), whereas reconnection close to or within the transition region may emphasize upflows (blue-wing enhancements) (e.g., \cite{2017MNRAS.464.1753H}). This height-dependent visibility provides an additional interpretive framework for the exclusive wing enhancements observed in some EEs.
The spectral evolution detailed in this work offers a set of observational constraints for future modeling efforts aimed at fully understanding the underlying drivers.}

{Our case studies allow us to propose a coherent physical interpretation:
\begin{enumerate}
	\item Bilateral enhancements are the direct spectroscopic signature of bi-directional reconnection outflows. Their persistence and the occasional appearance of high velocities ($\sim$100 $\mathrm{km}~\mathrm{s}^{-1}$) in late phases suggest ongoing or renewed energy release.
	\item Exclusive wing enhancements can be interpreted within two complementary frameworks. In the reconnection outflow model, they represent a geometric or temporal phase where one direction of the bidirectional jet dominates due to line-of-sight alignment or asymmetric heating. Alternatively, the rotating jet model suggests that such profiles could arise from a coherent jet structure undergoing strong spin, where a dominant rotational component or a localized torsional feature produces a single enhanced wing.
	\item The Evolutionary sequence captured across EE1--EE4 illustrates the life cycle of a reconnection event: from onset and dominant jetting (often blueshifted), through a phase of clear bidirectional flows, to a later phase where cooling and draining (redshifted components) become more prominent, sometimes accompanied by late-stage re-acceleration.
\end{enumerate}
}

{This work explores the nature of these EEs by linking their spectral evolution to the physics of magnetic reconnection. The observed events could be interpreted as tracing a sequence from plasma acceleration to subsequent heating and cooling processes. Future studies combining this spectroscopic analysis with high-resolution magnetograms to track photospheric magnetic cancellation at these sites will be valuable to test this interpretation and to refine the observational model.}

\begin{acknowledgments}
	This work was supported by the Strategic Priority Research Program of the Chinese Academy of Sciences (Grant No.\ XDB0560000), the National Natural Science Foundation of China (Grant Nos.\ 12325303, 11973084, 11803085, 12003064, U1831210, 11803002), the Yunnan Key Laboratory of Solar Physics and Space Science (Grant No.\ 202205AG070009), the Yunnan Provincial Science and Technology Department (Grant No.\ 202305AH340002), the Yunnan Fundamental Research Projects (Grant No.\ 202301AT070347 and 202301AT070349), and the Yunling Scholar Project of Yunnan Province. We also thank the NVST, IRIS, and SDO teams for providing high-cadence observational data.
\end{acknowledgments}

\bibliography{v2}{}
\bibliographystyle{aasjournalv7}

\end{document}